\documentclass[11pt,twoside]{article}
\usepackage{cspm-asp2006}
\usepackage{epsfig,graphicx,natbib,url}  
\usepackage{lscape}
\begin{document}

\markboth{S.~Wedemeyer-B\"ohm, O.~Steiner, J.~Bruls, and W.~Rammacher}{What 
          is heating the quiet-Sun chromosphere?}
\title{What is heating the quiet-Sun chromosphere?}
\author{S.~Wedemeyer-B\"ohm$^{1,2}$, O.~Steiner$^1$, J.~Bruls$^1$, and 
        W.~Rammacher$^1$}   
\affil {$^1$Kiepenheuer-Institut f\"ur Sonnenphysik, Sch\"oneckstr. 6,
        D-79104 Freiburg, Germany}    
\affil {$^2$Institute of Theoretical Astrophysics, University of Oslo,
        P.O. Box 1029 Blindern, N-0315 Oslo, Norway} 

\begin{abstract} 
  It is widely believed that the heating of the chromosphere in
  quiet-Sun internetwork regions is provided by dissipation of
  acoustic waves that are excited by the convective motions close to
  the top of the convection zone and in the photospheric overshoot
  layer.  This view lately became challenged by observations
  suggesting that the acoustic energy flux into the chromosphere is
  too low, by a factor of at least ten. Based on a comparison of TRACE
  data with synthetic image sequences for a three-dimensional
  simulation extending from the top layers of the convection zone to
  the middle chromosphere, we come to the contradicting conclusion
  that the acoustic flux in the model provides sufficient energy for
  heating the solar chromosphere of internetwork regions.  The role of
  a weak magnetic field and associated electric current sheets is also
  discussed.
\end{abstract}
\section{Introduction}
\label{sec:intro}

In two recent papers 
\citeauthor{2005Natur.435..919F} (2005, 2006, hereafter FC05 and FC06)
come to the conclusion that `high-frequency acoustic waves are not
sufficient to heat the solar chromosphere'. This conclusion is based
on a study of image sequences from the 1600\,\AA\ channel of TRACE
\textit{(Transition Region and Coronal Explorer)}.  The continuum
intensity at the wavelength of 160\,nm originates from the upper
photosphere.  There, acoustic waves suffer no longer strong radiative
damping and have not yet become strongly non-linear. Hence, the TRACE
images seem to be an excellent means for measuring the total
mechanical flux entering the chromosphere at its base.  However, the
mechanical flux cannot be measured directly but must be inferred from
the measured intensity fluctuation $(\Delta I/I)_{160\,\mathrm{nm}}$.
FC05 do this by comparing the average spectral power density for
$(\Delta I/I)_{160\,\mathrm{nm}}^\mathrm{obs}$ with those for
corresponding synthetic intensity fluctuations $(\Delta
I/I)_{160\,\mathrm{nm}}^\mathrm{synth}$ from simulations with the
one-dimensional radiation hydrodynamics code RADYN
(e.g., \cite{1997ApJ...481..500C}). 
They find that the power spectra fit best for simulation runs that
yield a mechanical flux at a height of 400\,km (integrated over 5 to
50\,mHz) of merely 438\,W\,m$^{-2}$ --- ten times less than is needed
to compensate the radiative losses of approximately 4300\,Wm$^{-2}$
(excluding Lyman\,$\alpha$) in the semi-empirical model of the
chromosphere by
\citeauthor{val81} (1981, hereafter VAL).
  
Here, we present new calculations based on the three-dimensional
radiation hydrodynamic simulation by
\citeauthor{wedemeyer04a} (2004, hereafter W04),    
which exhibit a dynamic pattern on a spatial scale too small to be
resolved with TRACE.  We discuss potential implications for the
heating of the solar chromosphere.

\section{Acoustic heating on small spatial scales} 
\label{sec:ah}

\subsection{Three-dimensional non-magnetic simulations}
\label{sec:sim}

The three-dimensional radiation hydrodynamics simulations by W04 
were carried out with the computer code 
\mbox{\textsf{CO$^\mathsf{5}$BOLD}}
(\cite{cobold})
and cover a horizontal region of 7\,\farcs7\,$\times$\,7\,\farcs7 and
extend from the upper convection zone at a depth of 1400\,km to the
middle chromosphere at a height of 1710\,km.  The spatial resolution
is 40\,km in the horizontal directions and 12\,km vertically in the
atmosphere.  Magnetic fields are not included so that this model
represents very quiet internetwork regions only.

In order to keep the computations feasible, the radiative transfer is
treated grey (i.e. frequency-independent) and under the assumption of
local thermodynamic equilibrium (LTE).  These are severe assumptions
for the upper photosphere and the chromosphere above but are
reasonable for the low photosphere in the framework of time-dependent
three-dimensional simulations.  The model chromosphere is -- in sharp
contrast to semi-empirical static models like, e.g., those by VAL --
highly intermittent in time and space, consisting of a dynamic
mesh-work like pattern of hot shock wave fronts and cool post-shock
regions in between.  The assumptions made for the radiative transfer
affect the temperature amplitudes and the chromospheric energy balance
but are reasonable for the low photosphere, where the acoustic waves
are excited.  The model provides sufficient acoustic energy to
counterbalance the radiative losses in the internetwork chromosphere
without need for further contributions from processes related to
magnetic fields
(\cite{wedemeyerphd}).

\begin{figure*}[tp] 
\centering 
  \includegraphics[width=0.95\textwidth]{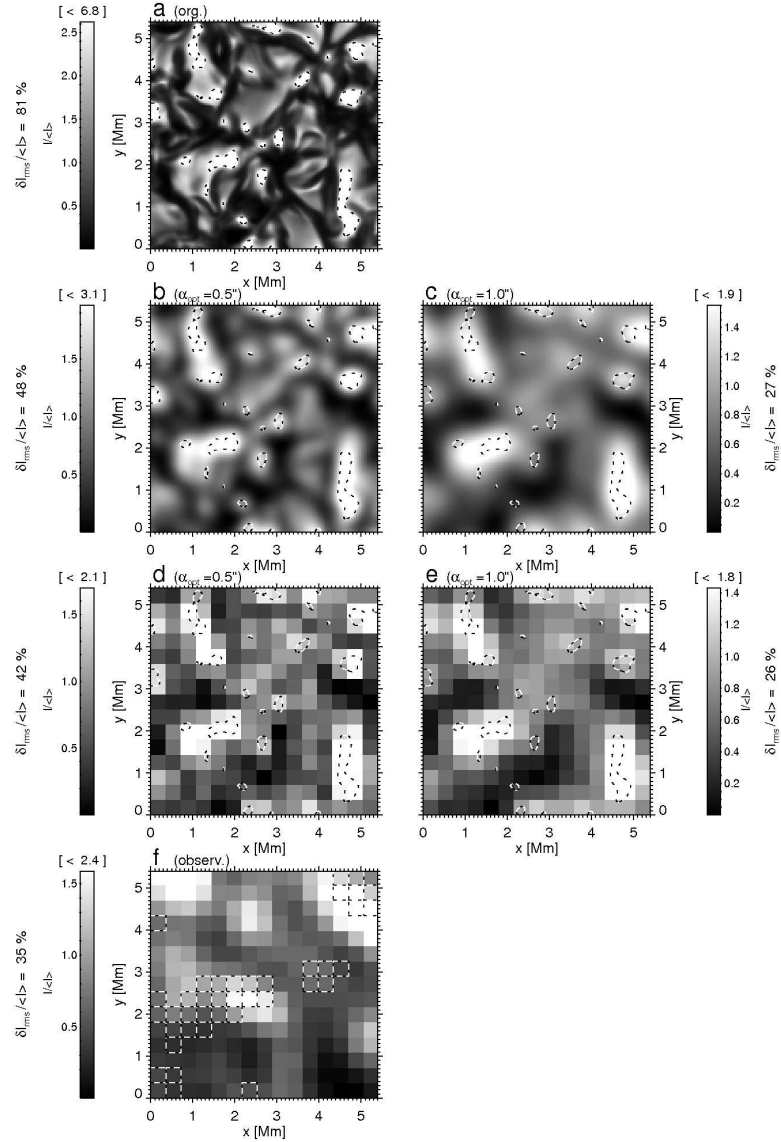}
  \caption{Synthetic intensity images for 160\,nm passband of TRACE:
    \textbf{a)}~original image, \textbf{b)}~degraded to a spatial
    resolution of 0\,\farcs5, and \textbf{d)}~after integration on
    TRACE 0\,\farcs5\ pixels.  The image degradation is repeated for a
    coarser resolution of 1\,\farcs0 in panel~\textbf{c)} but with the
    same pixel size in panel~\textbf{e)}.  The dashed contours for the
    original intensity in panel~a help to identify the enclosed
    regions in the other panels.  Panel~\textbf{f)} shows observed
    TRACE data from September 28$^\mathrm{th}$, 2005, taken from a
    data set kindly provided by Fossum \& Carlsson.  The dashed
    squares mark pixels that would be disregarded according to the
    pixel selection method by FC06.  The grey scale ranges of the
    panels are clipped individually. The maximum value is given in
    brackets above each legend. The intensity contrast is noted on the
    side of each panel. }
  \label{fig:intimg} 
\end{figure*} 

\subsection{Intensity synthesis}

The general procedure for our analysis can be summarised like this:
(i)~Synthesis of the continuum intensity at relevant ultra-violet (UV)
wavelengths, (ii)~integration of the intensity over wavelength to a
synthetic TRACE channel, (iii)~image degradation to the spatial
resolution due to the instrument's optics, (iv)~spatial intensity
integration for mimicking the TRACE detector (pixels), and finally
(v)~calculation of (average) power spectra from the intensity
sequences resulting from step (i) to (iv).

We use a short simulation sequence from W04 with a duration of 10\,min
and a cadence of 10\,s. For each snapshot we calculate continuum
intensity maps with the radiative transfer code
Linfor3D (\texttt{http://www.aip.de/}$\sim$\texttt{msteffen/}) at
wavelengths from 150\,nm to 180\,nm with an increment of 5\,nm,
covering the wavelength range relevant for the TRACE 160\,nm channel.
With increasing wavelength the intensity fluctuations become smaller, 
whereas the absolute intensity level grows strongly. Including longer
wavelengths thus effectively reduces the signal $I/<I>$ and the
resulting spectral power density.  The intensity sequences for
different wavelengths (150\,nm - 180\,nm) are now integrated according 
to the instrument's transmission curve
\citep[][see also  \texttt{http://www.lmsal.com/solarsoft}]{1999SoPh..187..229H},  
resulting in a short sequence for the synthetic TRACE channel. A 
typical intensity image is shown in Fig.~\ref{fig:intimg}a. 

In the next step the synthetic intensity images are convolved with a
point-spread function (PSF) in order to mimic the instrument's UV
optics (Fig.~\ref{fig:intimg}b).  Unfortunately, the PSF of TRACE is
poorly known but might be represented best by a Gaussian kernel of
width (FWHM) larger than 0\,\farcs5 but not more than 1\arcsec\
\citep{krijger01, 2004Icar..168..249S,   
       schrijver-privcomm06}.
%
The detector of TRACE has a pixel-scale of 0.5\,\arcsec/px.  A
synthetic pixel thus corresponds to spatial integration over (square)
regions of 9\,$\times$\,9 model grid cells (Fig.~\ref{fig:intimg}c).
Note that FC05 additionally apply a binning of $2 \times 2$ pixels but
FC06 use no binning.  An example of observational data from FC06 is
shown in Fig.~\ref{fig:intimg}d for comparison.

Figure~\ref{fig:intimg} clearly demonstrates that the pattern on small
spatial scales is lost and only the brightest features are still
visible in the synthetic image.  Consequently the intensity contrast
decreases from 81\,\% to 42\,\% for the displayed snapshot (89\,\% and
33\,\%, respectively, for the whole sequence).  The lower intensity
contrast of the observational data (35\,\% for the displayed region,
30\,\% for the whole time series for that region, and 26\,\% for the
time series of all pixels according to the mask by FC06) suggests that
the true resolution of TRACE might be coarser than the 0\,\farcs5
shown in Fig.~\ref{fig:intimg}c, which represents a lower limit only.
A resolution of 1\,\arcsec, on the other hand, might indeed be
considered an upper limit as it reduces the contrast of the synthetic
images to 33\,\% for the displayed snapshot and only 26\,\% for the
whole time series.

\subsection{Spectral power density}

\begin{figure}[tp] 
  \centering 
  \includegraphics[width=0.8\textwidth]{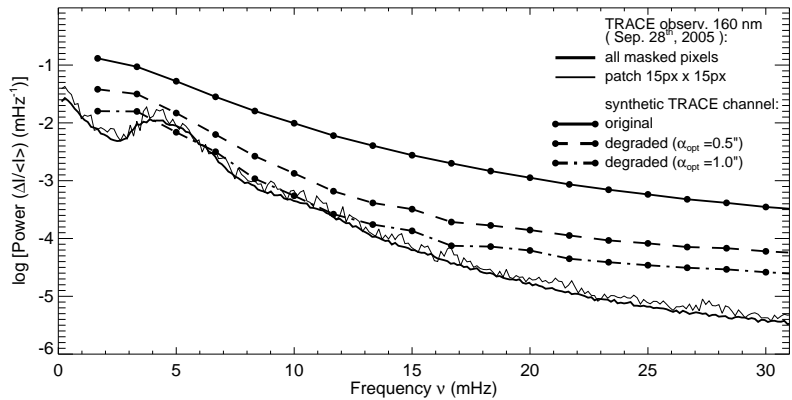}
  \caption{\label{fig:power}
    Average one-sided power density spectra for the short sequence of
    synthetic TRACE images (lines with dots) for original spatial
    resolution (solid), degradation to 0\,\farcs5 (dashed) and
    1\arcsec\ (dot-dashed) without pixel-binning. The solid dots mark
    the data points.  The two other curves are calculated from
    observational data from September 28$^\mathrm{th}$, 2005, in the
    same way as done for the synthetic sequences. The thick curve
    includes all pixels according to the mask by FC06, whereas the
    thin one is for the region displayed in Fig.~\ref{fig:intimg}d
    only.
}
\end{figure}

We derive the temporal spectral power density (one-sided) for each
spatial position in the synthetic image sequences separately.  The
horizontally averaged power spectra are shown in Fig.~\ref{fig:power}
for the sequence with the original spatial resolution of the input
model and also for the two sequences degraded to lower resolution
(0\,\farcs5 and 1\,\arcsec\,). For comparison we plot the power
spectra that we calculate from the observed TRACE sequence from
September 28$^\mathrm{th}$, 2005 (FC05), (i) for the pixel sample of
FC06 and (ii) for the small region shown in Fig.~\ref{fig:intimg}d.
Image degradation obviously suppresses power on small spatial scales.
The average power is reduced by a factor of 10 at 12\,mHz for the
degraded sequence with a resolution of 0\,\farcs5, with respect to the
original image sequence, whereas a resolution of 1\arcsec\ results in
a factor of 27 for the same frequency (see Fig.~\ref{fig:power}).  The
reduction in power increases with frequency up to $\sim 20$\,mHz as
higher frequencies are connected to smaller wavelengths (of acoustic
waves) that are more severely affected by a limited resolution than
large-scale fluctuations.  The power when using a PSF of 1\arcsec\
width agrees well with the observational results in the range 3\,mHz
to $\sim 12$\,mHz.

The discrepancy at low frequencies is due to the short sequence
duration, which results in a frequency resolution (see solid circles)
that is too small to resolve the local minimum at $\nu \approx
2.5$\,mHz. A longer synthetic image sequence at 160\,nm with a
duration of 76\,min and a cadence of 30\,s, that we calculated in
addition, shows results similar to those for the short sequence at
160\,nm with respect to absolute magnitude of power and the relative
effect of degrading the spatial resolution. The longer duration
provides much finer frequency sampling so that the average power
spectrum indeed exhibits a local minimum at low frequencies.


The increasing difference at higher frequencies is most likely 
connected to the assumption of LTE that had to be made in order
to keep the calculations computationally tractable. The assumption is
validated by comparison with a more detailed calculation with the RH code 
(see, e.g., \cite{uitenbroek00b})
for eight 2D slices from the model for a wavelength of 160\,nm, taking
into account deviations from LTE.  The rms fluctuations of the
resulting intensity samples are 150.8\,\% for Linfor3D (LTE),
199.3\,\% for RH in LTE, and 134.9\,\% for RH in NLTE.  The difference
for the LTE samples must be attributed to (i)~the assumption of 2D for
our application of RH in contrast to full 3D with Linfor3D,
(ii)~differences in opacity data (in particular concerning
photoionisation cross-sections of silicon), and (iii)~the neglect of
continuum scattering in Linfor3D.  The latter, however, is expected to
be of minor importance as there are few continuum scattering processes
that are relevant at wavelengths around 160\,nm. More important is the
correct treatment of bound-free transitions of silicon (taken into
account in RH).  They can be regarded as bound-free "two-level"
resonance scatterers and effectively smooth out fluctuations.
Nevertheless this qualitative comparison demonstrates that neglecting
non-LTE effects tends to produce too large relative intensity
fluctuations and with that an excess in average power for higher
frequencies as visible in Fig.~\ref{fig:power}.

In addition, a simulation sequence with non-grey radiative transfer 
(\cite{wedemeyerphd})  
was used for intensity synthesis at 160\,nm.  The resulting power
spectra at original and lowered resolution are very similar to the
grey sequence used here.

\section{Ohmic dissipation of chromospheric current sheets}
\label{sec:ohm}

\begin{figure}[]
  \centering
  \includegraphics[width=\textwidth]{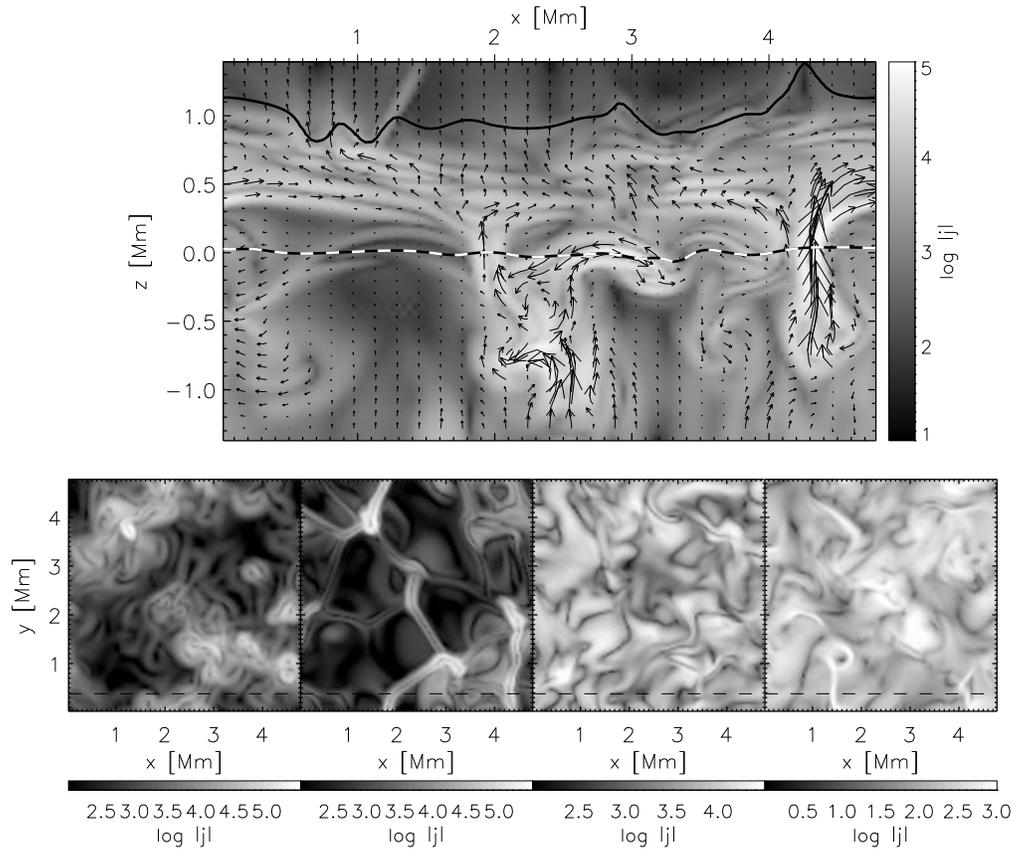}
  \caption{Logarithmic current density, $\log |j|$, in a vertical cross 
   section (top panel) and in four horizontal cross sections in a depth of 
   1180\,km below, and at heights of 90\,km, 610\,km, and 1310\,km above the 
   mean surface of optical depth unity from left to right, respectively.
   The arrows in the top panel indicate the magnetic field strength and
   direction. The dashed line indicates the position of the vertical section.
   [$j$] = $3\,\cdot\,10^5$ A/m$^{2}$.
  \label{fig:jg}}
\end{figure}

\subsection{Three-dimensional MHD simulation}

We have also carried out a three-dimensional simulation including
magnetic fields. The computational domain extends from a depth of
1400\,km below the mean surface of optical depth unity to 1400\,km
above it well into the chromospheric layers. The horizontal extent is
only slightly smaller than for the model by W04. See
\citeauthor{schaffenberger+al05} (2005, 2006) 
for details.
 
The MHD simulation starts with a homogeneous, vertical, unipolar
magnetic field of a flux density of 10\,G superposed on a previously
computed, relaxed model of thermal convection. This flux density is
thought to represent magneto-convection in a very quiet network-cell
interior. The magnetic field is constrained to have vanishing
horizontal components at the top and bottom boundary but lines of
force can freely move in the horizontal direction. Because the top
boundary is located in the chromosphere, the magnetic field can freely
expand with height through the photospheric layers into the more or
less homogeneous chromospheric field, different from conventional
simulations that extend to a height of typically 600\,km only.

A very common phenomenon in this simulation is the formation of a
`magnetic canopy field' that extends in a more or less horizontal
direction over expanding granules and between photospheric flux
concentrations. The formation of such canopy fields proceeds by the
action of the expanding flow above granule centres. This flow
transports `shells' of horizontal magnetic field to the upper
photosphere and lower chromosphere, where shells of different field
directions may be pushed close together, leading to a complicated
network of current sheets in a height range from approximately 400 to
900\,km.

This network can be seen in Fig.~\ref{fig:jg} (top), which shows, for
a typical snapshot of the simulation, the logarithmic current density,
$\log |j|$, together with arrows indicating the magnetic field
strength and direction. Figure~\ref{fig:jg} (bottom) shows from left to
right $\log |j|$ in four horizontal cross sections in a depth of
1180\,km below, and at heights of 90\,km, 610\,km, and 1310\,km above
the mean surface of optical depth unity. Higher up in the chromosphere
(rightmost panel), thin current sheets form along shock fronts, e.g.,
in the lower left corner near \mbox{$x = 1.4$\,Mm}.

\subsection{Ohmic dissipation}

In this simulation we have not taken an explicit magnetic diffusion
into account so that the effective electrical conductivity is
determined by the inherent artificial diffusion of the numerical
scheme. Therefore, due to lack of a realistic electric conductivity,
we here use molecular values when computing the ohmic dissipation of
the chromospheric current sheets.  Although the values might be orders
of magnitude too high, they still can be employed for the following
calculation, which gives a rough idea about the significance of ohmic
dissipation for chromospheric heating.
The typical current density in the height range from 400 to 900\,km of
Fig.~\ref{fig:jg} (top) is $j = 0.03$\,A\,m$^{-2}$. The electrical
conductivity, $\sigma$, in the photosphere and the lower chromosphere is
about 10 to 100\,A/Vm \citep{stix_book2}.  Using this value, the ohmic
dissipation is
\begin{equation}
P_j = \frac{1}{\sigma} j^2 \approx \frac{1}{10\ldots 100}\, 0.03^2 \approx
         10^{-5}\ldots 10^{-4}\, \mbox{W\,m}^{-3}\;.
\end{equation}
When integrating over a height range of 500\,km this heat deposition
leads to an energy flux of 5 to 50\,W\,m$^{-2}$. This value is about
two orders of magnitude short of being relevant for chromospheric
heating. However, the amplitude and width of the current sheets in the
simulation is determined by the spatial resolution of the
computational grid rather than by the molecular conductivity. The
effective conductivity in the simulation may easily be two orders of
magnitude lower so that magnetic heating by ohmic dissipation must be
seriously taken into account as a chromospheric heating agent. Further
simulations, taking explicit ohmic diffusion into account will clarify
this issue.

\section{Discussion}
\label{sec:discus}

%
The photosphere of the model by W04 can account for many 
aspects of observations of internetwork regions 
(see, e.g., \cite{leenaarts05}).  
Here we provide evidence that the UV intensity in the TRACE 160\,nm
channel is matched reasonably well, too.  When degraded to the spatial
resolution of the instrument, the synthetic intensity images presented
here have similar contrasts and also result in average spectral power
densities similar to the TRACE observations.  The power spectra
exhibit an enhancement around periods of 5\,min and otherwise a
decrease with frequency as it was found by FC05 and FC06 both from
observations and their 1D simulations.

%
At higher frequencies, however, we derive too much power from the
synthetic image series. This discrepancy is most likely due to the LTE
assumption made for the intensity synthesis with Linfor3D.  The test
calculations with the RH code show that taking into account deviations
from LTE reduces the intensity contrast by $\sim 60$\,\% compared to
the LTE case for a wavelength of 160\,nm.  The deviation from LTE at
the shorter wavelengths is mostly due to the strong over-ionisation of
Si\,I, which affects the continuum forming bound-free transitions.
Similar to true continuum scattering, this effect smoothes out
fluctuations, so that the LTE assumption used here actually tends to
produce too large fluctuations and thus an excess in power. We argue
that the derived reduction of $\sim 60$\,\% in intensity contrast can be
considered as an upper limit for the influence of this scattering-like
process. In contrast, the deviations from LTE are much smaller for the
longer wavelengths.  For a rough estimate we derive an average ratio
of LTE and NLTE intensities based on the RH calculations and use it to
scale the LTE intensities of the whole synthetic 160\,nm sequence.
This ``pseudo-NLTE'' time series leads to a further reduction in power
by approximately a factor of two around periods close to 5\,min to
four for higher frequencies.  The tendency of larger power reduction
with increasing frequency might be even more enhanced by non-LTE
effects being more important on time scales too short for relaxation
to equilibrium conditions.  The ``pseudo-NLTE'' estimate is of course
an order of magnitude approximation only and needs to be replaced with
detailed NLTE calculations that include not only 160\,nm but the whole
wavelength range relevant for the TRACE channel.
Taking into account line blanketing acts in the opposite direction by
reducing the intensity at the longer wavelengths and thus increasing
the relative fluctuations.  A comparison of the spectra calculated
with MULTI for the VAL model~A results in a reduction of integrated
intensity by a factor of $\sim 2$
(\cite{carlsson-privcomm06}).
But whether it really counterbalances the expected NLTE effects and if
so by how much and for which frequencies remains subject to detailed
radiative transfer calculations.

%
The computational box of the hydrodynamical simulation is small
(5.6\,Mm $\times$ 5.6\,Mm).  Consequently, the number of possible
oscillation modes is much smaller than in the real Sun so that the
power in the model is distributed over a smaller number of modes.  The
total energy content of the modes, however, remains the same.  Also
the finite grid resolution limits the treatment of high-frequency
waves but FC06 state that the neglect of frequency beyond 20\,mHz is
not an important omission.

%
When computing the mechanical energy flux from the intensity
fluctuations, FC05 and FC06 assume that the resolution of TRACE is
sufficient to fully capture the energy carrying waves.  FC06 state
that tests with 2D models related to the hydrodynamic simulations by 
\citeauthor{2004IAUS..223..385H} (2004) 
imply that only a factor of two in acoustic flux is missed due to
insufficient spatial resolution.  In contrast our 3D study presented
here shows that the unresolved small-scale pattern provides more power
by an order of magnitude or even more.

%
At this stage we cannot exactly determine how much of the additional
power hidden on small spatial scales is indeed transported to the
chromosphere and how much remains in the photosphere and therefore
does not contribute to chromospheric heating.  Separating the acoustic
flux from the total mechanical flux is not easy for our 3D model the
latter gets on average even negative in the middle photosphere as a
consequence of convective overshooting.  There is a local maximum of
the radiative flux divergence in the low chromosphere and a mechanical
counterpart of opposite sign that we interpret as evidence for the
conversion of mechanical energy provided by acoustic waves into
thermal energy, which is subsequently emitted in form of radiation
(\cite{wedemeyerphd}).
Integration of the mechanical flux divergence over the height range
from 570\,km to 1710\,km results in an acoustic energy flux of
$7.9\,\mathrm{kW\,m}^{-2}$ which is sufficient to
counterbalance the (semi-)empirically determined radiative energy
losses from the chromosphere (see, e.g, VAL).  Moreover, the total
mechanical flux at a height of 800\,km (which approaches the pure
accoustic flux there) amounts to
$3.8\,\mathrm{kW\,m}^{-2}$ and is thus in line with the 1D models by
\citeauthor{2005ccmf.confE..60R} (2005). 

On the other hand, the energy flux due to Ohmic dissipation of current
sheets in the MHD model chromosphere is only of the order of 5 to
50\,W\,m$^{-2}$. This value must be considered as a lower limit only
because of the uncertainty in the electrical conductivity.  It can
thus not be ruled out that the magnetic heating contribution might be
underestimated by even two orders of magnitude.

Although the energy balance of the model chromosphere is affected by a
too simple treatment of the radiative transfer, the calculations
presented here suggest that the acoustic contribution to chromospheric
heating cannot be neglected. Magnetic fields may also play an indirect
role for guiding acoustic waves and for mode conversion
(\cite{suematsu90, ulmschneider91, jefferies06}).

\section{Conclusions}
\label{sec:conc}

Based on the analysis presented here we conclude that TRACE might miss
acoustic power by at least one order of magnitude due to its limited
spatial resolution.  Horizontal flows, as they are clearly seen
within the reversed granulation pattern, contribute to the intensity
variations although their relevance for the energy transport to the
upper layers is not clear yet.  

Matching the empirical power spectra with a model that has enough
acoustic power to counterbalance the observed chromospheric emission
should not be interpreted such that high-frequency waves alone are
sufficient to heat the chromosphere.  It is not clear from this study
how important the small relative power contribution of high-frequency
waves is for chromospheric heating. Moreover, the model used here
exhibits a chromosphere with pronounced dynamics and a highly
inhomogeneous structure -- in sharp contrast to static and
one-dimensional semi-empirical VAL-like models. The ubiquitous shock
waves, which produce our model chromosphere's characteristic
structure, play certainly an important role for the energy balance of
our model chromosphere.

It must be emphasised that -- owing to simplifications that are
necessary in order to keep the problem computationally tractable --
the topic should be addressed again with detailed NLTE calculations.
Nevertheless, the presented study clearly demonstrates that
three-dimensional radiation hydrodynamics simulations in principle can
reproduce the TRACE observations of internetwork regions without
strong need for magnetic fields, although weak fields are most likely
an integral part of those regions.  We conclude that an acoustic
``basal flux''
\citep{schrijver87} 
must still be considered a possible option for heating the
chromosphere of internetwork regions.

\acknowledgements We thank the CSPM organisers for a very good meeting
and M.~Carlsson, K.~Muglach, C.~Schrijver, R.~Rutten, V.~Hansteen, and
H.~Uitenbroek for discussion and helpful comments.  We are grateful to
A.~Fossum for kindly providing observational data and useful remarks
regarding data reduction.  Thanks go also to M.~Steffen, H.-G.~Ludwig,
and B.~Freytag for their support for Linfor3D and \mbox{CO$^5$BOLD}.
H.~Uitenbroek friendly provided his RH~code.  This work was supported
by the {\em Deutsche Forschungs\-gemein\-schaft (DFG)}, grant Ste~615/5.


\end{document}